\documentclass[aps,prl,twocolumn,superscriptaddress]{revtex4-2}

\pdfoutput=1

\usepackage{amsmath,amsfonts}
\usepackage{graphicx}
\usepackage[colorlinks=true]{hyperref}

\begin{document}

\title{Lasing in a ZnO waveguide: clear evidence of polaritonic gain obtained by monitoring the continuous exciton screening.}

\author{Geoffrey Kreyder}
\affiliation{Institut Pascal, Université Clermont Auvergne, CNRS, Clermont INP, 63000 Clermont-Ferrand, France}
\author{Léa Hermet}
\affiliation{Institut Pascal, Université Clermont Auvergne, CNRS, Clermont INP, 63000 Clermont-Ferrand, France}
\author{Pierre Disseix}
\affiliation{Institut Pascal, Université Clermont Auvergne, CNRS, Clermont INP, 63000 Clermont-Ferrand, France}
\author{François Médard}
\email[]{Francois.MEDARD@uca.fr}
\affiliation{Institut Pascal, Université Clermont Auvergne, CNRS, Clermont INP, 63000 Clermont-Ferrand, France}
\author{Martine Mihailovic}
\affiliation{Institut Pascal, Université Clermont Auvergne, CNRS, Clermont INP, 63000 Clermont-Ferrand, France}
\author{François Réveret}
\affiliation{Institut Pascal, Université Clermont Auvergne, CNRS, Clermont INP, 63000 Clermont-Ferrand, France}
\author{Sophie Bouchoule}
\affiliation{Centre de Nanosciences et de Nanotechnologies, CNRS, Université Paris-Saclay, 91120 Palaiseau, France}
\author{Christiane Deparis}
\affiliation{Université Côte d’Azur, CNRS, CRHEA, rue Bernard Gregory, Sophia Antipolis, 06560 Valbonne, France}
\author{Jesús Zuñiga-Pérez}
\affiliation{Université Côte d’Azur, CNRS, CRHEA, rue Bernard Gregory, Sophia Antipolis, 06560 Valbonne, France}
\affiliation{Majulab, International Research Laboratory IRL 3654, CNRS, Université Côte d’Azur, Sorbonne Université, National University of Singapore, Nanyang Technological University, Singapore, Singapore}
\author{Joël Leymarie}
\affiliation{Institut Pascal, Université Clermont Auvergne, CNRS, Clermont INP, 63000 Clermont-Ferrand, France}

\date{\today}

\begin{abstract}
The coherent emission of exciton-polaritons was proposed as a means of lowering the lasing threshold because it does not require the dissociation of excitons to obtain an electron-hole plasma, as in a classical semiconductor laser based on population inversion. In this work we propose a method to prove clearly the polaritonic nature of lasing by combining experimental measurements with a model accounting for the permittivity change as a function of the carrier density. To do so we use angle resolved photoluminescence to observe the lasing at cryogenic temperature from a polariton mode in a zinc oxide waveguide structure, and to monitor the continuous shift of the polaritonic dispersion towards a photonic dispersion as the optical intensity of the pump is increased (up to 20 times the one at threshold). This shift is reproduced thanks to a model taking into account the reduction of the oscillator strength and the renormalization of the bandgap due to the screening of the electrostatic interaction between electrons and holes. Furthermore, the measurement of the carriers lifetime at optical intensities in the order of those at which the polariton lasing occurs enables us to estimate the carrier density, confirming that it is lower than the corresponding Mott density for zinc oxide reported in the literature.
\end{abstract}


\maketitle

\section{Introduction}

First introduced by Hopfield \cite{Hopfield1958} and Agranovich \cite{Agranovich1960} to describe the light propagation in bulk semiconductors, exciton-polaritons have been extensively investigated since 1992 in vertical microcavities \cite{Weisbuch1992}. Exciton-polaritons, hereafter referred to as polaritons, result from the strong coupling of photons and excitons, which are electron-hole pairs in Coulombic interaction. Contrary to the particles involved in an electron-hole plasma, polaritons are bosons and when their density exceeds a critical value, they condensate into a single state which emits coherent light, resulting in polariton lasing \cite{Imamoglu1996}. The working mechanism is not the stimulated emission of photons as in conventional lasers, but the stimulated polariton scattering induced by the final state occupation \cite{Bajoni2012}. This process is predicted to occur at a much lower particle density \cite{Johne2008} with a smaller gain length than the half cavity length required in a conventional laser \cite{Souissi2022}. This has promoted polariton lasers as promising candidates for low-power coherent sources compatible with large packing densities and, thus, they are particularly adapted to the field of optical interconnections where a drastic reduction of the energy-consumption is required \cite{Hill2014}. However, the maximum operation temperature of a polariton laser depends on the exciton stability, which is material dependent. Zinc oxide (ZnO) is a wide-bandgap semiconductor with stable and robust excitons for which polariton lasing has been demonstrated up to room temperature \cite{Li2013}. It is thus an attractive material for realising polariton lasers, provided that the carrier density is kept below the Mott density, which corresponds to the dissociation of all excitons into unbound electrons and holes \cite{Haug2004}.\\

Waveguides, which confine photons through total internal reflection, represent a geometry for polariton lasers alternative to the most standard vertical one \cite{Walker2013}. Polaritonic waveguides combine numerous advantages: (i) inherent very low radiative losses, given that they operate at wave vectors outside the light cone, (ii) a significant spatial overlap between the excitonic medium and the electromagnetic field, and (iii) an eased fabrication, as they consist only of a semiconductor slab embedded between two optical cladding layers. Strong coupling in waveguides has been first demonstrated in structures based on J-aggregates \cite{Ellenbogen2011} and gallium arsenide \cite{Walker2013}. However, these materials are only suitable for low power or low temperature applications. Recently, other groups have reported promising results for future devices based on guided polaritons working at room temperature: polariton lasing and amplification in ZnO \cite{Jamadi2018}, long-range coherent polariton condensate flow in a halide perovskite waveguide \cite{Su2018}, ultrafast polariton modulation with GaN quantum wells \cite{DiPaola2021} and also polariton lasing in GaN-based circular guides \cite{Delphan2022}.\\

In this work we propose a method to clearly evidence the polaritonic nature of lasing. To do so we have measured accurately the continuous shift of the polariton dispersion in a ZnO-based waveguide at cryogenic temperature for different pumping intensities, from below lasing threshold to far above lasing threshold. The experimental results are confronted with a modelling of the permittivity accounting for the progressive screening of bound electron-hole pairs. As the optical intensity of the pump is increased, the laser energy moves from the polaritonic dispersion, whose shape is mostly controlled by excitonic contributions to the dielectric function, towards the dispersion of the photonic waveguide mode (for intensities 20 times larger than the polariton lasing threshold). This observation is explained because of the screening of the electrostatic Coulomb interaction, leading to a reduction of the oscillator strength and a bandgap renormalisation, thus modifying the complex permittivity of the guiding medium. The observed shift of the dispersion to smaller wave vectors, as predicted by our model, enables us to determine a reduction by 22\% of the oscillator strength at the polariton lasing threshold. We also perform time-resolved photoluminescence to determine the decay time of carriers allowing us to calculate a carrier density at threshold of $\mathrm{3 \times 10^{17} \, cm^{-3}}$. A careful comparison with the Mott density value obtained from the literature confirms the existence of polaritons at such a density.\\

\section{Sample structure and optical setup}

\begin{figure}
	\includegraphics[width=\columnwidth]{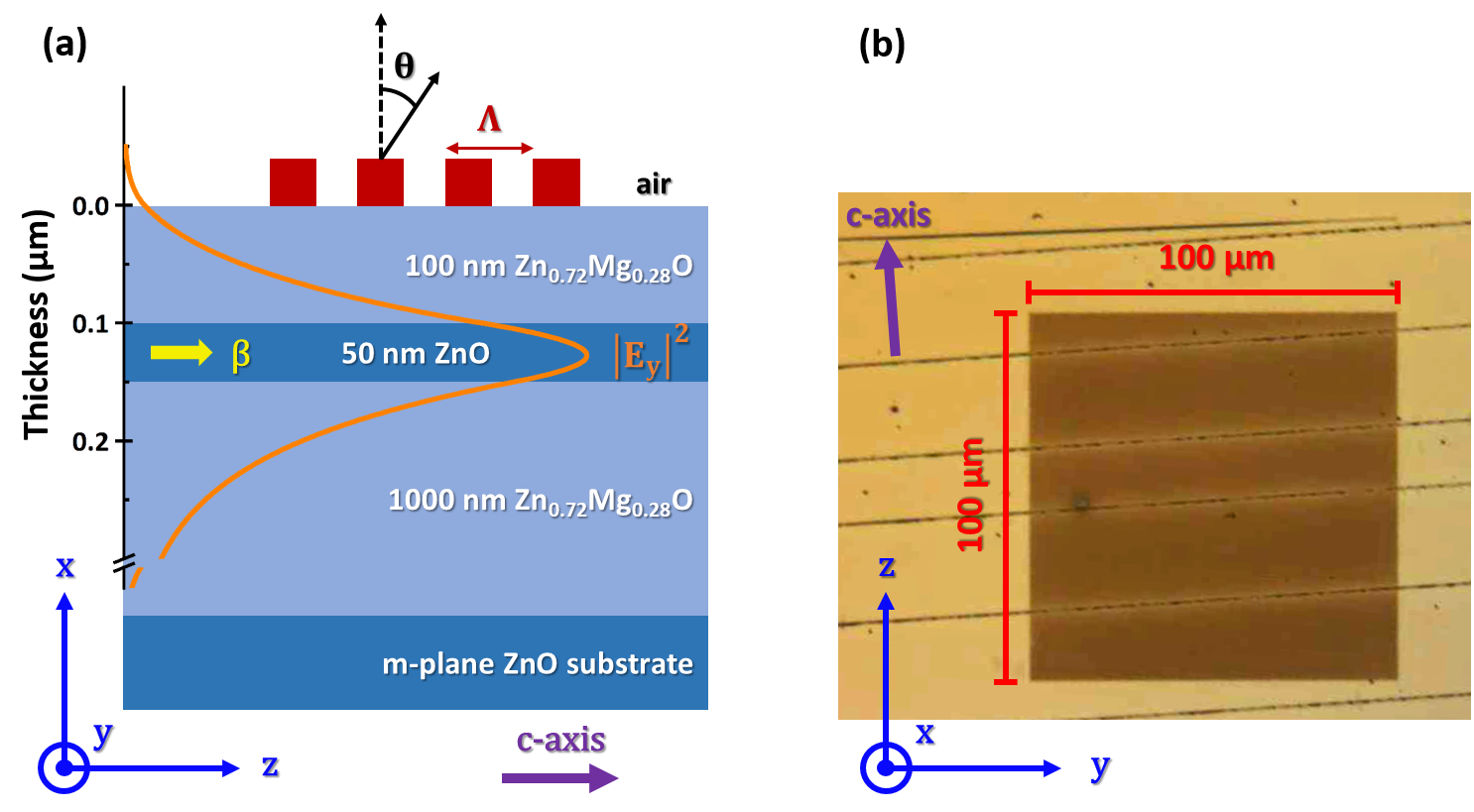}
	\caption{\label{1}(a) Sketch of the sample structure with the definition of axes used in the calculation of guided modes, the emission angle $\theta$ at the surface and the period $\Lambda$ of the grating. The ZnO guiding layer is 50 nm thick with the propagation constant $\beta$ depicted in yellow. The square modulus of the electric field for the TE0 mode is plotted in orange. (b) Image of the sample surface obtained using an optical microscope showing a $\mathrm{100 \times 100 \, \mu{}m^2}$ diffraction grating (dark square) and cracks perpendicular to the c-axis (purple arrow) as darker lines. The twist angle between the grating and the in-plane c-axis is smaller than $\mathrm{4^{\circ}}$ leading to a negligible error on the wave vector measurement.}
\end{figure}

The structure investigated in the present study consists of a ZnO based optical waveguide (50 nm thick) grown by molecular beam epitaxy on a $\mathrm{Zn_{0.72}Mg_{0.28}O}$ buffer layer (1000 nm) deposited onto an m-plane ZnO substrate \cite{Zuniga-Perez2016}. A $\mathrm{Zn_{0.72}Mg_{0.28}O}$ upper cladding (100 nm) was added to prevent surface recombinations and to obtain a better confinement of the electromagnetic field. A sketch of the structure is displayed in Fig.\ref{1}a. During the epitaxial growth, cracks are naturally induced by the mismatch of both the lattice parameters and the thermal expansion coefficients of ZnO and $\mathrm{Zn_{0.72}Mg_{0.28}O}$. Cracks form perpendicular to the in-plane c-axis of the crystal and are irregularly spaced: in the current sample the typical distance between two cracks lies between 5 to 40 µm typically (see the optical microscopy image of the sample in Fig.\ref{1}b. Due to the refractive index contrast between ZnO and air, cracks can act as mirrors and thus two parallel cracks form a Fabry-Perot resonator. As displayed on Fig.\ref{1}b, a $\mathrm{SiO_2}$ grating ($\mathrm{100 \times 100 \, \mu{}m^2}$) with a $\Lambda$=190 nm period was elaborated by electron beam lithography on a hydrogen silsesquioxane (HSQ) resin on top of the structure to enable the outcoupling of the guided waves. Due to the difficulties to perfectly align the mask along the crystallographic axes, there is a small twist angle between the grating and the c-axis of about $4^\circ$. Consequently, strictly speaking we measure the projection of the wave vector along z, but this results in a negligible variation.\\

The Bragg’s diffraction law relates the emission angle $\theta$ outside the sample to the propagation constant $\beta$ of the mode:
\begin{equation}
	\beta = \frac{2\pi}{\lambda_0} \, sin(\theta) - m \, \frac{2\pi}{\Lambda} \quad \mathrm{with} \quad m \in \mathbb{Z}
\end{equation}

The dispersion of the guided polaritons is thus measured by Fourier-space spectroscopy combined with a real-space filtering. The emission of the sample was collected through a 100x NUV microscope objective with a 0.5 numerical aperture and its Fourier plane (focal plane) was imaged onto the spectrometer slit by two spherical lenses. The spectrometer has a 1 m focal length and a 1200 grooves per mm grating to obtain a resolution better than 0.05 nm. For spatial filtering a pinhole was placed in the focal plane of the first lens; the pinhole diameter of 0.8 mm corresponds to an investigated surface of 200 $\mathrm{\mu{}m^2}$ on the sample. This spatial filtering enabled us to select the light emitted only from a small area of the grating coupler deposited on the sample but also to avoid, as much as possible, collecting light from adjacent cavities. A Q-Switched laser emitting at 266 nm with a 400 ps pulse duration and 20.6 kHz repetition rate was used for the excitation. We have used a side excitation, focusing the laser through an aspherical lens with a spot surface of 1220 $\mathrm{\mu{}m^2}$.\\

For the time resolved photoluminescence (TRPL), the waveguide is excited using the third harmonic of a titanium-sapphire laser with a pulse duration of 150 fs. After being spectrally dispersed by a 600 groove per mm grating, the luminescence is temporally analysed using a streak-camera.

\section{Numerical calculation of guided modes dispersion and permittivity model to account for varying carrier density}

In order to analyse the experimental results, guided modes in the waveguide are calculated by solving Maxwell’s equations with proper boundary conditions. In the region between two cracks, the structure is invariant by translation along y and z axes (see Fig.\ref{1}a) and the propagation equation for transverse electric (TE) polarisation modes writes:
\begin{equation}
	\begin{split}
	\frac{d^2 E_y (x)}{dx^2}+(k_0^2 \epsilon_r-\beta^2 ) \, E_y (x)=0  \\ \mathrm{with} \quad \vec{E}=E_y (x) \,  e^{(i(\omega{}t-\beta{}z))} \,  \vec{u}_y
	\end{split}
\end{equation}
where $k_0$ is the vacuum wave vector and $\beta$ the propagation constant in the waveguide which becomes complex when losses are taken into account through the permittivity $\epsilon_r$. The continuity of transverse components of the electric and magnetic fields at each interface leads to an implicit equation whose resolution is carried out in the complex plane to determine the propagation constant $\beta$ as a function of the energy E. We limit our study to the TE0 mode, for which the electric field confinement is the strongest.\\

The complex permittivity of ZnO is based on the value measured by spectrometric ellipsometry at 300 K \cite{Schmidt2003,Schmidt2007}. We have modelled the excitonic transitions with harmonic oscillators and removed them to obtain a background value $\epsilon_b$ that retains only the contribution of both band-to-band and higher energies transitions. At low temperature, the background permittivity is blueshifted by 70 meV corresponding to the bandgap change from 300 K to 5 K \cite{Meyer2004}. Then, the permittivity can be expressed using the transition energy $E_j$, the broadening $\gamma_j$ and the oscillator strength $f_j$ of ZnO excitons as follows:
\begin{equation}
	\epsilon(E) = \epsilon_b (E) + \sum_{j=A,B} \frac{f_j}{E_j^2-E^2+i\gamma_j E}
\end{equation}

\begin{figure}
	\includegraphics[width=\columnwidth]{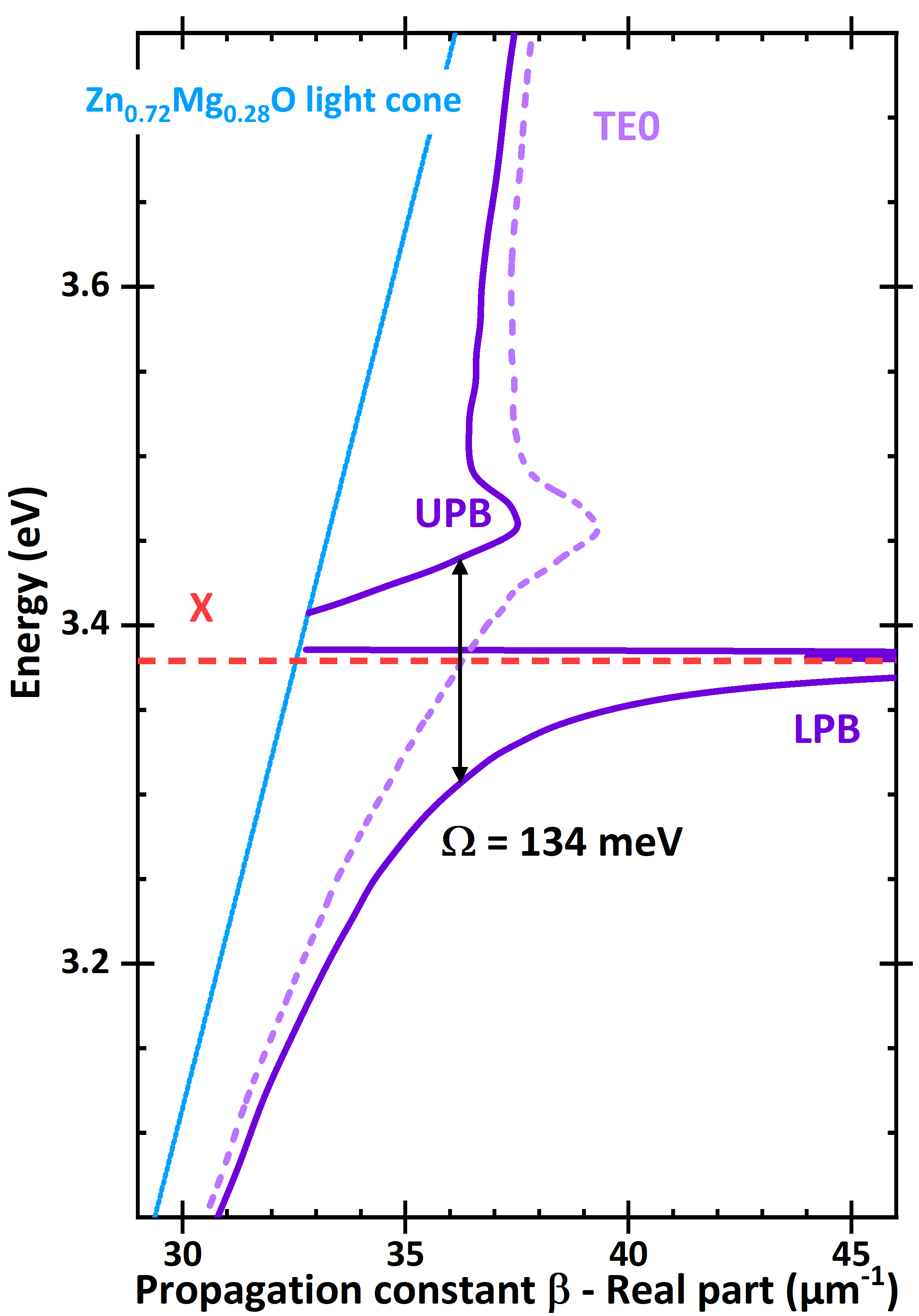}
	\caption{\label{2}Numerical calculation of the guided polariton dispersion in purple showing a Rabi splitting of 134 meV between the lower and upper branches. The TE0 photonic mode (light purple dashed line) is obtained using only the background permittivity of ZnO without excitonic contributions. The $\mathrm{Zn_{0.72}Mg_{0.28}O}$ cone (light blue) corresponds to a linear dispersion with a constant refractive index of 1.9. The red dotted line indicates the mean energy of the excitonic transitions.}
\end{figure}

At low temperature and low density, we used excitonic parameters from the work by Mallet et al \cite{Mallet2013}. On Fig.\ref{2}, the light cone of $\mathrm{Zn_{0.72}Mg_{0.28}O}$ is also represented in light blue using a constant optical index of 1.9 \cite{Teng2000} and the photonic TE0 mode (purple dashed line) is determined when the excitonic contribution is removed ($f_j$ = 0) from the permittivity. An example of the calculated dispersion curve for the waveguide at a temperature of 5 K is displayed as a purple line in Fig.\ref{2}, evidencing clearly the polaritonic nature of the guided mode with lower and upper polariton branches being separated by a Rabi splitting $\Omega$ of 134 meV. This value is smaller than the previously published one \cite{Jamadi2018} as the model used in our first work coupled a linear dispersion of the TE0 mode (with a constant effective refractive index) with a harmonic oscillator. It has been demonstrated in reference \cite{Brimont2020} that a coupled oscillator model strongly overestimates the deviation of the polaritonic dispersion from the photonic one. Moreover, this is a crude model as it does not account for the dispersive nature of the refractive index and its variation as a function of the carrier density. Thus a clear demonstration of the polaritonic nature of the mode, in the absence of the upper polariton branch, can only be obtained by monitoring the shift of its dispersion from a polaritonic one to a photonic one, following thereby the Coulomb potential screening of excitons up to the Mott transition.\\

As the carrier density increases, the Coulomb interaction between electrons and holes is progressively screened leading to the vanishing of excitons at the Mott density. Both experimental and theoretical results in bulk semiconductors have shown that the exciton energy does not depend on the carrier density \cite{Manzke1998,Banyai1986}. This is explained by a compensation between the reduction of the exciton binding energy $\mathrm{E_b}$ and the shrinkage of the bandgap. We propose to model the effect of the carrier density using only one parameter $\eta$, which represents the percentage of oscillator strength remaining for a given carrier density n:
\begin{equation}
	\eta=\frac{f(n)}{f(n=0)}
\end{equation}

As both the oscillator strength and the binding energy are linked to the exciton Bohr radius, we can express the permittivity as a function of $\eta$:
\begin{equation}
	\begin{split}
	\epsilon(E)=\epsilon_b \left(E-E_b \times \left[1-\eta^{1⁄3}\right]\right) \\+ \sum_{j=A,B} \frac{\eta \, f_j}{E_j^2-E^2+i\gamma_j E}
	\end{split}
\end{equation}

The first term of the equation accounts for the shrinkage of the bandgap and, thus, the spectral of the background contribution, while the second term concerns the reduction of the amplitude of the excitonic transition due to the screening of excitons by carriers. The term $[1-\eta^{1⁄3}]$ comes from the fact that the excitonic binding energy is proportional to $a_B^{-1}$ where $a_B$ is the excitonic Bohr radius, while the oscillator strength is proportional to $a_B^{-3}$ \cite{Elliott1957}.

\section{Experimental measurement of the polariton dispersion as a function of optical pumping intensity}

\begin{figure}
	\includegraphics[width=\columnwidth]{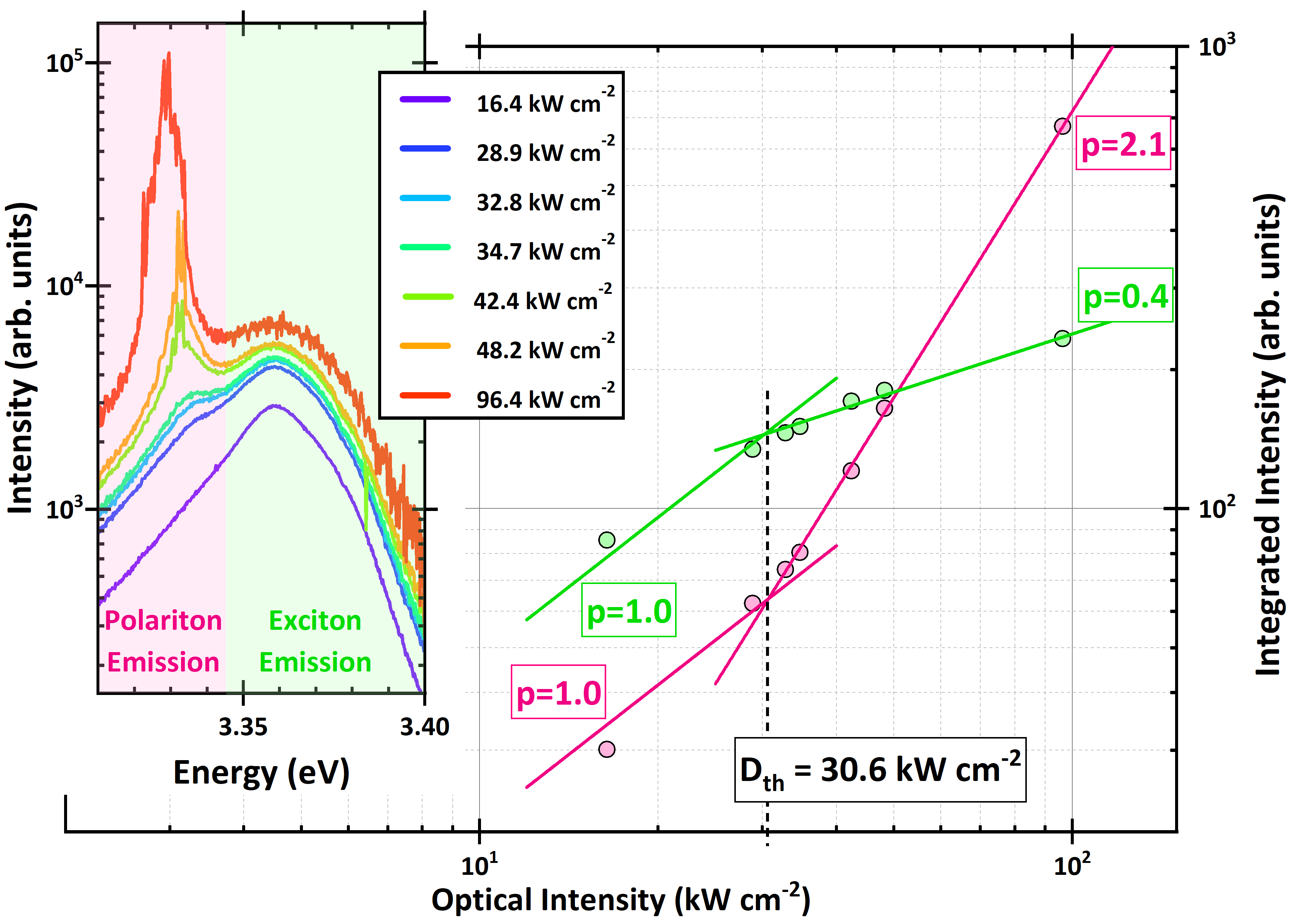}
	\caption{\label{3}Photoluminescence spectra recorded at low temperature ($\mathrm{T_{sample} \, = \, 5 K}$) for optical intensities increasing from 16.4 $\mathrm{kW.cm^{-2}}$ to 96.4 $\mathrm{kW.cm^{-2}}$ (left panel); the integrated intensity is plotted as a function of the optical intensity of the pump power (right panel) for the polaritonic emission (pink, 3.31 to 3.345 eV) and for excitonic transitions (green, 3.345 to 3.40 eV) ; results are shown on a log-log scale to evidence the superlinear increase for polaritons (p=2.1) and the supralinear one for excitons (p=0.4) above the threshold at $\mathrm{D_th}$ = 30.6 $\mathrm{kW.cm^{-2}}$.}
\end{figure}

The photoluminescence of the ZnO waveguide has been recorded at low temperature ($\mathrm{T_{sample} \, = \, 5 \, K}$). The spectra are shown on the left panel of Fig.\ref{3} for various optical intensities. For the lowest intensity ($\mathrm{D \, = \, 16.4 \, kW.cm^{-2}}$), the spectrum is dominated by a broad emission centered at 3.36 eV corresponding to the emission of both donor-bound and free excitons. The broadening is due to strain fluctuation in the structure as further discussed below. As the optical intensity is increased, another peak appears at lower energy, which progressively resolves into a comb of sharp lines: these lines are successive modes of the Fabry-Pérot cavity formed by two consecutive cracks. Their intensity increases clearly faster than the excitonic emission. To prove unambiguously the nonlinearity of the emission, we have plotted on the right side of Fig.\ref{3} the integrated intensity which is proportional to the light emitted, versus the optical pumping intensity D. Above a threshold at $\mathrm{D_{th} \, = \, 30.6 \, kW.cm^{-2}}$, the area of the lower energy peak (integrated between 3.31 and 3.345 eV, pink) increases as a power law with an exponent of 2.1, whereas the excitonic emission (integrated between 3.345 and 3.340 eV, green) increases only as $\mathrm{D^{0.4}}$. Indeed, the emission attributed to polaritons is superlinear while the excitonic one is supralinear, associated to an excitonic reservoir depletion that will be considered in the discussion section. Note that an extensive study of lasing in the same structure, with more experimental points, has already been published in reference \cite{Jamadi2018}.\\

\begin{figure}
	\includegraphics[width=\columnwidth]{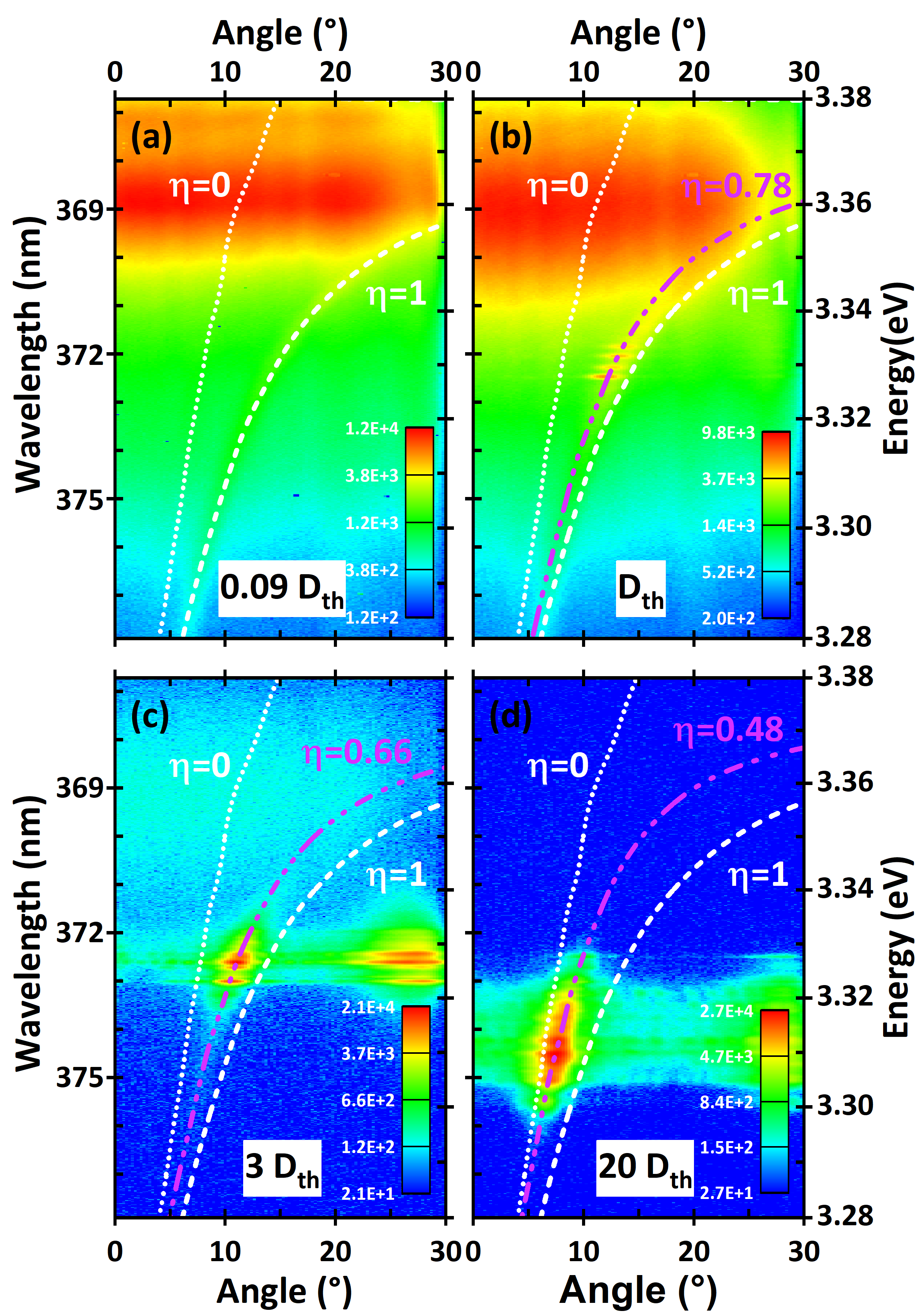}
	\caption{\label{4}Experimental dispersion curves of the polaritonic guided mode at $\mathrm{T_{sample} = 5 K}$ as a function of the optical intensity of the excitation laser. Calculated dispersions are also plotted using a variable parameter $\eta$ (percentage of oscillator strength) accounting for the exciton screening. The white dashed curve correspond to $\eta$=1, the white dotted one to $\eta$=0 and the magenta lines are best fits to the experimental data. Each panel corresponds to a different optical intensity: (a) 2.9 $\mathrm{kW.cm^{-2}}$, (b) 31.8 $\mathrm{kW.cm^{-2}}$ corresponding to lasing threshold, (c) 95.4 $\mathrm{kW.cm^{-2}}$ and (d) 636 $\mathrm{kW.cm^{-2}}$.}
\end{figure}

The dispersion relation of the ZnO waveguide is also investigated experimentally at the same cryogenic temperature ($\mathrm{T_{sample} = 5 K}$) for optical intensities ranging from 2.9 $\mathrm{kW.cm^{-2}}$ to 636 $\mathrm{kW.cm^{-2}}$. The emission is recorded as a function of the wavelength in ordinates and, thanks to the dispersion grating, with an angular resolution in abscissa. The results are shown in a false colours logarithmic scale on Fig.\ref{4} for four selected intensities. On each panel, the calculated dispersions are also plotted using various percentage of oscillator strength: $\eta$=1 corresponds to the low-density regime (white dashed line), $\eta$=0 accounts for the permittivity at the Mott density (white dashed dot line) and the best fit to the experimental dispersion using a variable $\eta$ is depicted in light purple.\\

For the lowest optical intensity (Fig.\ref{4}a – 2.9 $\mathrm{kW.cm^{-2}}$), a strong dispersionless emission is observed at energies between 3.35 eV and 3.38 eV. It corresponds to the photoluminescence of uncoupled excitons whose emission cannot be guided in the waveguide and are thus emitted vertically. This luminescence is mainly dominated by two broad peaks at 3.3765 eV and 3.3605 eV: they are attributed respectively to free excitons and neutral donor bound excitons, in agreement with reference \cite{Meyer2004} for unstrained ZnO. The significant broadening originates from large strain variations in the area between the cracks, as we spatially integrate the emission from excitons emitting close and far from the cracks. The presence of excitons confirms that the carrier density is well below the Mott density. At lower energy, a dispersive feature corresponds to the emission of the guided polaritons. The white dashed line is the best fit of the experimental data: it allows us to fix the physical parameters of the structure assuming the oscillator strength is not affected at the lowest excitation ($\eta$=1). The dispersion relation of the waveguide is also reported at the Mott density ($\eta$=0, dotted curve) for comparison.\\

As the excitation is increased (Fig.\ref{4}b – 31.8 $\mathrm{kW.cm^{-2}}$, \ref{4}c – 95.4 $\mathrm{kW.cm^{-2}}$ and \ref{4}d – 636 $\mathrm{kW.cm^{-2}}$), the dispersion curve shifts continuously towards the low emission angle, corresponding to lower propagation constants for a given energy. This is due to screening effects as the charge carrier density is increased: bandgap renormalization and oscillator strength reduction both lead to a lower permittivity value, which blueshifts the guided mode dispersion (at constant wavevector). This behaviour is well accounted for by our model if we decrease the percentage of oscillator strength $\eta$ from 1 to 0.78 (Fig.\ref{4}b), 0.66 (Fig.\ref{4}c) and 0.48 (Fig.\ref{4}d) as it can be seen from the agreement between the fit (purple lines) and the experimental data.\\

Another interesting phenomenon appears on Fig.\ref{4}b, where discrete peaks appear superimposed on top of the polaritonic dispersion: they correspond to the resonances of a Fabry-Pérot cavity formed by two cracks that become visible as the gain overcomes the losses. This is a clear evidence of lasing which it is corroborated by the nonlinear increase of the integrated intensity as a function of the excitation shown on Fig.\ref{3}. The peaks are aligned on the calculated dispersion corresponding to $\eta$=0.78: this constitutes a conclusive proof of the polaritonic nature of the lasing mechanism. The nonlinear emission is still visible at higher power densities (Fig.\ref{4}c and \ref{4}d) but the peaks become broader as other adjacent cavities with different sizes contribute to the collected emission. Simultaneously, the energy of the lasing modes decreases, as predicted by theoretical calculations based on semi-classical Boltzmann equations by Solnyshkov et al \cite{Solnyshkov2014}. This behaviour can be explained by an enhanced polariton relaxation along the lower polariton branch due to increased polariton-polariton interactions.  At 636 $\mathrm{kW.cm^{-2}}$, an optical pumping intensity corresponding to 20 times the one at threshold, the dispersion curve comes even closer to the $\eta$=0 curve (dotted white line). The uncertainty on the permittivity used in the calculations as well as the smaller energy difference between the photonic and the polaritonic dispersions at this energy, do not allow us to conclude any longer whether the lasing is based on polaritons or on an electron-hole plasma. We have to mention that on the last two figures, a blurred emission is detected at high angles and away from any polaritonic dispersion (at angles between $\mathrm{20^\circ}$ and $\mathrm{30^\circ}$): this is due to the diffraction of the undispersive emission from the cracks on the pinhole used to spatially select the emission.

\section{Discussion on the lasing regime}

\begin{figure}
	\includegraphics[width=\columnwidth]{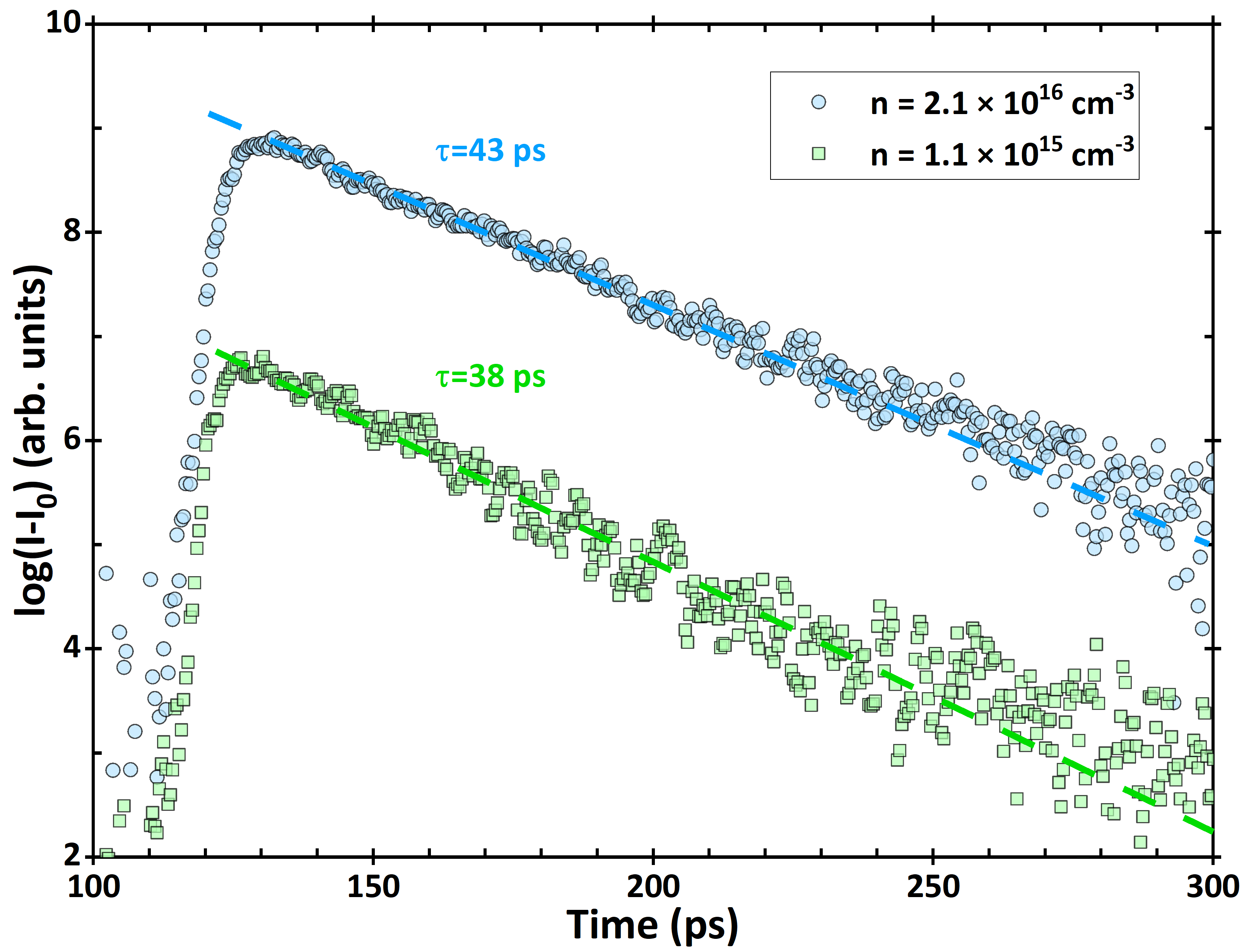}
	\caption{\label{5}Time resolved photoluminescence of the free excitonic transition (3.3765 $\pm$ 0.0025 eV), and eventually ionized electron-hole pairs (energy up to 3.41 eV), for a carrier density n=$\mathrm{2.3 \times 10^{15} \, cm^{-3}}$ (green squares) and n=$\mathrm{4.7 \times 10^{16} \, cm^{-3}}$ (blue circles). The ordinate axis corresponds to the natural logarithm of the intensity to evidence the mono-exponential decay. The green and blue dotted lines are guides for the eyes.}
\end{figure}

The polariton laser is fed from a reservoir of uncoupled excitons with large wave vectors. We propose to determine now the density of carriers n in this reservoir. To do so, we have measured the decay time of the excitonic transitions that do not couple to the waveguide and that can be recorded in the luminescence spectrum emitted from the surface. Time resolved measurements are shown on Fig.\ref{5} for two different optical intensities corresponding to n in the range of $\mathrm{10^{15}}$ to $\mathrm{5 \times 10^{16} \, cm^{-3}}$, as detailed below. To obtain the decay time of carriers in the reservoir (both free excitons and eventually ionized electron hole pairs, as explained later), we have measured the photoluminescence decay of the waveguide with a temporal resolution smaller than 5 ps and integrated over an energy range from 3.3765 eV (free A exciton energy from Fig.\ref{4}a) up to 3.41 eV (3 times $\mathrm{k_BT_e}$ with $\mathrm{T_e \sim 130 K}$ the excitonic temperature at lasing threshold). We have plotted the natural logarithm of the luminescence intensity I subtracted from the background $\mathrm{I_0}$ to unambiguously prove the monoexponential decay. The lifetime of the carriers appears not to depend on the carrier density and we will consequently use a mean value $\tau$=40 ps for calculations.\\

The carrier density can be easily determined for these measurements using the following formula, where the laser pulse is assumed to be a Dirac delta function:
\begin{equation}
	n=\frac{E_{pulse}}{E_{laser}} \, \times \, \frac{\alpha_{ZnO}}{S}
\end{equation}

We define $E_{pulse}$ the energy per pulse transmitted to the ZnO layer, $\mathrm{S \, = \,2000 \, \mu{}m^2}$ the surface of the spot and $E_{laser}$ = 4.66 eV the energy of the excitation laser. The part of the photons absorbed by the $\mathrm{Zn_{0.72}Mg_{0.28}O}$ cladding is taken into account using $\mathrm{\alpha_{ZnMgO} \, = \, 1.1 \, \times \, 10^5 \, cm^{-1}}$ \cite{Teng2000} as well as the transmission coefficients of the interfaces between each layer. The generated carriers are assumed to be not transferred from the $\mathrm{Zn_{0.72}Mg_{0.28}O}$ cladding to the ZnO guide as they are mostly trapped into alloy fluctuations at low temperature. As the ZnO thickness is less than the optical absorption length \cite{Klingshirn2007}, the relevant thickness to consider for the evaluation of n is 1/$\mathrm{\alpha_{ZnO}}$ with $\mathrm{\alpha_{ZnO}=1.9 \, \times \, 10^5 \, cm^{-1}}$ \cite{Muth1999}. Using the above formula, we can estimate that the carrier density is varying from $\mathrm{2.3 \times 10^{15} \, cm^{-3}}$ for the green curve in Fig.\ref{5} to $\mathrm{4.7 \times 10^{16} \, cm^{-3}}$ for the blue one.\\

Coming back to the dispersion measurement of Fig.\ref{4}, the pulse duration of the laser (400 ps) is large enough compared to the carrier lifetime (40 ps) and we can use the quasi-continuous approximation. The carrier density is now written as:
\begin{equation}
	n=\frac{D}{E_{laser}} \, \times \, \tau \, \times \, \alpha_{ZnO}
\end{equation}

We can thus deduce a carrier density $\mathrm{n_{th}=3 \, \times \, 10^{17} \, cm^{-3}}$ for an optical pumping intensity $\mathrm{D \, = \, 30.6 \, kW.cm^{-2}}$ corresponding to the polariton lasing threshold. This value has to be compared to the Mott density $\mathrm{n_{Mott}}$ from the literature. Using the static screening model developed by Versteegh et al in reference \cite{Versteegh2011}, we were able to calculate $\mathrm{n_{Mott}=7 \, \times \, 10^{17} \, cm^{-3}}$ for an excitonic temperature of 130K, which is determined under the high excitation conditions. This value, compared to $\mathrm{n_th}$, further confirms the polaritonic nature of the laser.\\

As the carrier density is increased (see Fig.\ref{4}c and \ref{4}d), we expect to reach the Mott density; however, this phenomenon is not clearly observed, even at 636 $\mathrm{kW.cm^{-2}}$. One possible process explaining this, at least partially, is the depletion of the excitonic reservoir by the stimulated scattering of polaritons to the final state, its population being clamped above the polariton lasing threshold \cite{Wouters2007}. Indeed, we have demonstrated on Fig.\ref{3} that the excitonic emission increases only with an exponent 0.4 above threshold. Thus, a pumping intensity 20 times larger than the one at threshold should lead to an increase of the carrier density by only a factor of 3.3. The optical intensity of 636 $\mathrm{kW.cm^{-2}}$ would thus correspond to a carrier density of $\mathrm{10 \, \times \, 10^{17} \, cm^{-3}}$, close to the Mott transition in zinc oxide. This estimation is consistent with our inability to discriminate experimentally, at this large pumping intensity, the nature of the lasing gain (Fig.\ref{4}d).

\section{Conclusion}

We have studied at cryogenic temperature a ZnO-based waveguide that supported guided polariton modes. Thanks to outcoupling gratings deposited on the surface of the structure, the dispersion curve of the TE0 mode has been experimentally measured using Fourier space spectroscopy to prove its polaritonic nature. The numerical resolution of Maxwell equations allows to compare the measured dispersions with a purely photonic one, where the excitonic contributions to the permittivity have been precisely removed, showing a significant difference. As the power density of the optical excitation is increased, a nonlinear emission is observed at a threshold of 30.6 $\mathrm{kW.cm^{-2}}$, with a small shift of the dispersion. A model has been developed to account for carrier screening effects in the excitonic reservoir through the reduction of the oscillator strength and bandgap renormalization. It demonstrates that the lasing is of polaritonic nature as the oscillator strength is only reduced by 22\% at threshold. The lifetime of carriers ($\tau$=40 ps) in the reservoir has been determined through time resolved photoluminescence and it is used to compute the carrier density at threshold. The value of $\mathrm{n_{th}=3 \, \times \, 10^{17} \, cm^{-3}}$ is found to be lower than the Mott density reported in the literature, corroborating that the emission originates from polaritons.\\

The complex permittivity model accounting for excitonic screening used in this work, combined to the measurement of the polaritonic dispersion as a function of the optical intensity, can be used for other polaritonic systems containing different materials, even at high temperatures. Our description of the permittivity may be improved with a more formal description of the interacting electron-hole gas as described in the literature. Doing so, we could also access to the whole dielectric function below and above the Mott transition to simulate the gain from the electron-hole plasma and to predict the transition to a classical laser.

\begin{acknowledgments}
	The authors acknowledge funding from the French National Research Agency (ANR-16-CE24-0021 PLUG-AND-BOSE)
\end{acknowledgments}

\bibliography{library}

\end{document}